\documentclass[a4paper]{article}

\usepackage{INTERSPEECH2020}
\usepackage{threeparttable}
\usepackage{hyperref}

\title{Towards Learning a Universal Non-Semantic Representation of Speech}
\name{Joel Shor$^1$, Aren Jansen$^2$, Ronnie Maor$^1$, Oran Lang$^1$, Omry Tuval$^1$, F\'elix de Chaumont Quitry$^3$, Marco Tagliasacchi$^3$, Ira Shavitt$^1$, Dotan Emanuel$^1$, Yinnon Haviv$^1$}

\address{
    \footnotesize
  $^1$Google Research, Israel\\
  $^2$Google Research, Mountain View\\
  $^3$Google Research, Zurich}
\email{joelshor@google.com}

\begin{document}

\maketitle

\begin{abstract}
The ultimate goal of transfer learning is to reduce labeled data requirements by exploiting a pre-existing embedding model trained for different datasets or tasks.  The visual and language communities have established benchmarks to compare embeddings, but the speech community has yet to do so. This paper proposes a benchmark for comparing speech representations on non-semantic tasks, and proposes a representation based on an unsupervised triplet-loss objective. The proposed representation outperforms other representations on the benchmark, and even exceeds state-of-the-art performance on a number of transfer learning tasks. The embedding is trained on a publicly available dataset, and it is tested on a variety of low-resource downstream tasks, including personalization tasks and medical domain. The benchmark\footnote[4]{\url{https://www.tensorflow.org/datasets/catalog/overview\#audio}}, models\footnote[5]{\url{https://tfhub.dev/s?q=nonsemantic-speech-benchmark}}, and evaluation code\footnote[6]{\url{https://github.com/google-research/google-research/tree/master/non\_semantic\_speech\_benchmark}} are publicly released.
\end{abstract}


\section{Introduction}
One of the most powerful uses of deep learning is finding a good representation for a given domain. Despite progress on representations in the visual domain \cite{zhai2019visual} and the language domain \cite{wang2018glue}, no such universal representation exists for the speech domain. One reason is a lack of standard benchmark tasks to compare different methods; for example, existing speech representations tend to focus on one problem set such as speaker recognition or speech emotion recognition \cite{latif2020deep}. In this paper, we propose a set of benchmark speech tasks that are diverse, to require that ``good'' representations contain general speech information, and targeted, to allow good performance when compared with task-specific representations.

We propose a specific set of publicly available tasks, called the ``NOn-Semantic Speech benchmark'' (NOSS), to assess the general usefulness of speech representations on ``non-semantic'' tasks. What we call ``non-semantic'' tasks do not include tasks like automatic speech recognition and phone classification, which require sub-second granularity. They do include paralinguistic tasks such as speech emotion recognition, as well as tasks such as speaker identification, language identification, and medical diagnosis. In addition, we introduce a new set of intra-speaker sub-tasks from existing tasks where a model is trained and evaluated on speech from a single speaker. These intra-speaker tasks help measure which representations are most useful for personalization, which is an increasingly-relevant use-case as more computation is performed on-device.

A good speech representation should be high-performing on a diverse set of downstream tasks using simple models. In addition, it should be useful in transfer learning with small amounts of data for a new task. This use-case is relevant for model personalization, such as user-specific emotion recognition or speaker identification. 


We introduce a representation, TRILL (TRIpLet Loss network), which is learned in a self-supervised manner on speech containing clips from AudioSet \cite{gemmeke2017audio}. Using the techniques of \cite{triplet2017}, the network represents audio such that segments which are closer in time are also closer in the embedding space. We demonstrate that this simple proxy objective is highly effective in learning a strong representation for multiple non-semantic speech tasks.

We evaluate TRILL and other representations on our benchmark by training small models built on top of the representations and comparing their performances. In addition, we explore transfer learning by fine-tuning TRILL using data from the downstream tasks. This is an advantage of learned representations over non-learned ones. Pre-training via transfer learning can sometimes outperform models trained on a single dataset \cite{zhai2019visual}, and this is also the case in our benchmark. Using transfer learning, we are able to achieve a new state-of-the-art in many of the tasks, surpassing previously published results which sometimes were hand-crafted for those specific datasets.

In summary, our contributions are:
\begin{enumerate}
    \item We define a new benchmark for comparing representations on non-semantic speech tasks using previously published data. In addition, we add a sub-category of personalization tasks.
    \item We demonstrate that a single representation learned in an unsupervised manners performs best on this benchmark. We compare it to existing representations, feature-based and learned.
    \item We fine-tune our best performing representation, further boosting results. This method sets a new state-of-the-art on many previously published tasks.
    \item We distill our learned representation to a model that can run inference and training on-device, and we open-source the original and distilled models.
\end{enumerate}

\section{Background}
\begin{table*}[t]
\centering
\caption{Datasets for downstream benchmark tasks.}
\begin{tabular}{ c c c c c c c} \toprule
  Dataset        & Has          & Target & Number of     & Number of     & Number of &  Average  \\ 
                 & intraspeaker &  & classes & samples & speakers & duration (secs) 
  \\
  \midrule
 *VoxCeleb~\cite{nagrani2017voxceleb}        & No       &  Speaker identification   &1,251           & 153,514           & 1,251      & 8.2 \\
 VoxForge~\cite{maclean2018voxforge}       & No       & Language identification   & 6             & 176,438           & 13,559   & 5.8 \\
 Speech Commands\cite{warden2018speech} & Yes      & Command                   & 12            & 105,829           & 2,618     & 1.0 \\
 CREMA-D~\cite{cao2014crema}        & Yes      & Emotion                   & 6             & 7,442             & 91      & 2.5 \\
 SAVEE~\cite{haq2009speaker}          & Yes      & Emotion                   & 7             & 480               & 4     & 3.8 \\
 DementiaBank~\cite{boller2005dementiabank}    & No       &  Dementia/healthy         & 2             & 210               & 210       & 70.0 \\
 \bottomrule
\end{tabular}
*Results in our study used a subset of Voxceleb filtered according to YouTube's privacy guidelines. The numbers here represent the full dataset. TensorFlow Datasets and our open source evaluation framework use the full dataset.
\label{tab:noss}
\end{table*}

Transfer learning and domain adaptation have been extensively studied in machine learning \cite{pan2009survey}. Recent research has mostly focused on deep representation learning methods, either supervised, semi-supervised, or unsupervised. Successful representations improve the sample efficiency of ML algorithms by extracting most information out of the raw signal from the new data before any task-specific learning takes place. This strategy has been used successfully in many application domains \cite{tan2018survey, cheplygina2019not}. 

An important step in learning a good representation is having a standard benchmark to evaluate it. Such benchmarks should contain a variety of downstream tasks, representing different tasks in the domain. Such benchmarks have been developed in vision and NLP \cite{zhai2019visual, wang2018glue}. 

There are three standard approaches to adapting a representation to multiple, potentially heterogeneous, downstream tasks. One approach is to train a task-specific linear classifier on the embeddings produced by a pre-trained network, whose parameters are kept frozen~\cite{yosinski2014transfer}. A second approach is to fully fine-tune. Generally, fine-tuning matches or outperforms the performance of fully-supervised models trained on the downstream tasks~\cite{zhai2019visual, kong2019panns}, especially when the amount of labeled data is small. A third approach is multi-task learning. This has been applied in the speech domain \cite{Pascual2019}, although not on a wide range of tasks. It is usually favored when the downstream tasks are all applied on the same input set.

There are many methods for learning audio representations. The work in~\cite{kong2019panns} trained an embedding for audio classification on AudioSet. Other work has demonstrated the value of supervised~\cite{daredevil2016, Pascual2019}, semi-supervised \cite{Parthasarathy_2018}, or unsupervised representations. The unsupervised audio representation literature is especially diverse (L$^3$~\cite{L32017}, AuDeep~\cite{audeep2017}, Autoregressive Predictive Coding~\cite{apc2019}, Contrastive Predictive Coding~\cite{cpc2018}, metric learning~\cite{triplet2017}, autoencoding~\cite{latif2018variational}). However, these methods were evaluated on just one or a limited set of downstream tasks. 

In other domains, training a strong representation requires a very large, general dataset. AudioSet~\cite{gemmeke2017audio} is the largest dataset for general purpose audio machine learning, serving as an audio equivalent of ImageNet. Even when restricted to only the samples with speech tags, it surpasses all datasets in size and variability. It has been used to learn a general purpose audio embedding in \cite{kong2019panns}, and can be used for multiple speech tasks.

\section{Non-Semantic Speech Benchmark (NOSS)}
\label{sec:benchmark}

To standardize the assessment of non-semantic speech representations, we introduce NOSS. This section describes the benchmark in detail (summarized in Table~\ref{tab:noss}). These tasks reflect different properties of the speech signal, and they vary in size and difficulty. Personalization and on-device training is increasingly important, so we include ``Intra Speaker'' tasks for the datasets when applicable. Intra-speaker tasks are an important addition because they also test task adaptation for small amounts of data, and that representations do not just rely on speaker identity.

\textbf{Inter-speaker tasks} are described in Table~\ref{tab:noss}. They were chosen so that they span a wide variety of non-semantic tasks, including  ``paralinguistic'' tasks such as speech emotion recognition, as well as other kinds of tasks such as speaker identification, language identification, and one task of voice-based medical diagnosis. Each dataset has a different number of target classes, and different number of examples.

\label{sec:intra_speaker}
An important use-case of task adaptation is personalization---training and evaluating a model on data only of a specific person. We call these tasks \textbf{intra-speaker tasks}. The accuracy is averaged over all speakers. 
Note that most inter-speaker tasks divide the speakers into disjoint groups in the train/test split, and intra-speaker tasks use all speakers for both training and testing. Not all tasks have a meaningful intra-speaker versions; it is meaningless to train and test on the same speaker in tasks with labels that depend only on the speaker's identity, such as language identification, medical diagnosis, and speaker identification. The intra-speaker tasks are: CREMA-D, SAVEE, and Speech Commands.

\section{Experiments}

\subsection{TRILL representation}
\label{sec:triplet}

\begin{table*}
\begin{threeparttable}
\centering
\caption{Average performance of the different embeddings on a number of downstream tasks.}
\begin{tabular}{ c c c c c c c } \toprule
                            & VoxCeleb1*     & VoxForge      & SpeechCommands          & CREMA-D   & SAVEE & DementiaBank \\
\midrule\midrule
Prev SOTA                       & -         & 89.0\cite{revay2019multiclass} & 91.1 \cite{kagglesc2018}  & \textbf{74.0}$\dagger$ \cite{ghaleb2019} & 61.0\cite{haq2009speaker} & \textbf{80.6}$\ddagger$ \cite{db2017}   \\ 
\midrule  
Mel / MFCC                      & 12.2     & 79.2          & 47.7              & 52.8      & 55.6    & 67.7 \\
OpenSmile                       & 2.5       & 78.0          & 36.5              & 53.7        & 62.6    & 68.8 \\
Random Network                  & 11.0      & 73.0          & 42.0              & 52.1      & 48.6    & 67.9 \\
YAMNet top                      & 3.1       & 67.0          & 40.7              & 52.2      & 45.4    & 64.8 \\
YAMNet layer 10                 & 16.6          & 86.1          & 73.1              & 66.4      & 62.3   & 70.0 \\
VGGish top                      & 12.5       & 80.8          & 28.3              & 51.3       & 49.8   & 68.3 \\
VGGish FCN 1                    & 14.8      & 85.1          & 52.7              & 55.7      & 57.7   & 68.7 \\
TRILL top                       & 16.4       & 83.8          & 60.4              & 64.9      & 53.7   & 68.2 \\
\midrule
TRILL layer 19                  & 17.7       & 88.1          & 74.0              & 67.8       & 67.8  & 67.2 \\
TRILL layer 19, MobileNet 2048d & \textbf{17.9}       & 83.4          & 74.9              & 68.1       & 60.0  & 68.1 \\
\midrule
TRILL finetuned                 & 17.6       & \textbf{94.1} & \textbf{91.2}    & 69.5 & \textbf{68.6} & 73.1 \\
\midrule
\bottomrule
\end{tabular}
*Results in our study use a small subset of Voxceleb1 filtered according to YouTube's privacy guidelines. Interested readers can run our study on the full dataset using TensorFlow Datasets and our open-source evaluation code. State-of-the-art performance on the un-filtered dataset can be found in \cite{nagrani2017voxceleb}. 
$\dagger$ Also uses video features. $\ddagger$ Also uses text features.
\label{tab:results-inter}
\end{threeparttable}
\end{table*}

Non-semantic aspects of the speech signal (e.g., speaker identity, language, and
emotional state) generally change more slowly than the phonetic and lexical
aspects used to explicitly convey meaning.  Therefore, we expect a good
representation for non-semantic downstream tasks to be considerably more stable
in time than what is required for ASR applications. To take advantage of this intuition, we follow the work of~\cite{triplet2017} (Section 3.2.5) and use temporal proximity as a self-supervision signal. 

More formally, consider a large, unlabeled speech collection represented as
a sequence of spectrogram context windows $X = x_1 x_2 \dots x_N$, where each
$x_i \in \mathbb{R}^{F \times T}$.  Our goal is to
learn a map $g: \mathbb{R}^{F \times T} \rightarrow \mathbb{R}^d$ from
spectrogram context windows to $d$-dimensional space such that $\|g(x_{i}) -
g(x_{j})\| \leq \|g(x_{i})-g(x_{k})\|$ when $|i-j| \leq |i-k|$.  We can express
this desired relationship as a learning objective using triplet loss-based
metric learning as follows.  First, we sample from $X$ a large collection of
example triplets of the form $z = (x_i, x_j, x_k)$ (the so-called anchor, positive,
and negative examples), where $|i-j| \le \tau$ and $|i-k| > \tau$ for some
suitably chosen time scale $\tau$.  The loss incurred by each triplet is then
given by 

\begin{equation*}
  \mathcal{L}(z) = \sum_{i=1}^N \left[ \|g(x_i)\!-\!g(x_j)\|_2^2 -
    \|g(x_i)\!-\!g(x_k)\|_2^2 + \delta \right]_+,
  \label{eq:loss}
\end{equation*}

\noindent
where $\|\!\cdot\!\|_2$ is the $L_2$ norm, $[\cdot]_+$ is standard hinge loss,
and $\delta$ is a nonnegative margin hyperparameter. We use the
now-standard within-batch semi-hard negative mining
technique~\cite{schroff2015facenet}.


The TRILL model is trained on the subset of AudioSet training set clips possessing the speech label.  We set $\tau$ to 10 seconds, the maximum duration of each AudioSet clip. This makes the training task is primarily same clip / different clip discrimination.
Following \cite{daredevil2016,triplet2017}, we (i) take as input log mel
spectrogram context windows with $F=64$ mel bands and $T=96$ frames representing
0.96 s of input audio (STFT computed with 25 ms windows with step 10 ms); and
(ii) employ the Resnetish~\cite{daredevil2016} variant of the standard
ResNet-50 architecture followed by a $d=512$ dimensional embedding layer. Since the ResNet's final average pooling operation destroys the sub-second
temporal structure, we also consider representations defined by
earlier convolutional blocks.

\subsection{Other representations}
\label{sec:lrep}

We compared TRILL to several representations used frequently on non-semantic tasks. These include classical methods like Mel spectrograms and OpenSmile \cite{eyben2010opensmile}, which is the de-facto standard in emotion recognition\cite{latif2020deep}. For Mel spectrograms, we tested many different configurations and selected the best one on each task, and for OpenSmile we used the ComParE16 acoustic parameter set \cite{compare2016}. 

We also compared to existing learned representations. 
\emph{YAMNet}~\cite{plakal_ellis_2020} is a state-of-the-art network for audio classification \cite{kong2019panns}, trained on AudioSet. \emph{VGGish}~\cite{daredevil2016} is an audio embedding produced by training a modified VGGNet model to predict video-level tags from the Youtube-8M dataset. We also compared to our own network with random initialization, a technique which has been shown to produce good embeddings \cite{tian2019deep}.

\subsection{Experimental Method}
\label{experimental_method}

\subsubsection{Evaluation methods}

In order to evaluate the usefulness of the representations described in Section~\ref{sec:lrep}, we train small models to solve the downstream NOSS tasks (Section~\ref{sec:benchmark}). For each representation / task pair, we explore different downstream models, representation aggregation techniques, and normalization methods. 

We train shallow models using Scikit-Learn \cite{scikit-learn}. We experimented with logistic regression, random forest classifiers, and linear discriminator analysis (LDA) as the downstream models.  Despite the shallowness of the above models, we achieve competitive results with the best previously-reported results on some of the benchmark tasks. 

The embeddings were aggregated over time using average pooling, except for VoxCeleb which used NetVLAD aggregation \cite{ar2015netvlad}. Datasets which have multiple utterances per speaker were normalized for each speaker using $L^2$ normalization.  

Some of the downstream tasks have fixed canonical splits. For the inter-speaker tasks on those datasets, we report numbers on the canonical splits (Speech Commands and VoxCeleb). For the other datasets, and for all the intra-speaker tasks, we perform five random train / test splits and report the average. For intra-speaker tasks (Section \ref{sec:intra_speaker}), we train and test on one speaker at a time, then average results across splits and across speakers.

\subsubsection{Network layer and model distillation}

For the representations generated from pre-trained neural networks (TRILL, VGGish, YAMNet), we experimented with both the final output and two intermediate representations. For TRILL, we tried the final 512-dimensional embedding layer and the pre-ReLU output of the first 512-depth convolutional layer (subsequently referred to as layer 19 due to TensorFlow convention). We found that layer 19 performed best on our tasks. For VGGish, we tried the final layer, and the first fully-connected layer. For YAMNet, we test the final pre-logit layer, and the fifth depth-separable convolutional layer outputs. 

To make the network more useful, we used distilliation to train a smaller model of similar quality. We used a truncated MobileNet architecture~\cite{howard2017mobilenets} to predict the original TRILL network's layer 19 embeddings. This distillation reduces model parameters and multiplies by factors of 5.6X (9M $\rightarrow$ 1.6M) and 25X (1.5B $\rightarrow$ 59M), respectively.


\begin{table}
\centering
\caption{Intra-speaker task performance}
\begin{tabular}{ c c c c } \toprule
                                &            &       & Speech \\ 
                                &  CREMA-D   & SAVEE & Commands \\ 
\midrule
Mel / MFCC                      & 47.7      & 77.2  & 73.3 \\
OpenSmile                       & 47.6      & 67.8  & 72.2 \\
Random Network                  & 43.4      & 73.9  & 69.0 \\
YAMNet top                      & 39.1      & 64.4  & 70.3 \\
YAMNet layer 10                 & 50.8      & 79.2  & \textbf{77.6} \\
VGGish top                      & 43.1      & 68.7  & 70.2 \\
VGGish FCN 1                    & 50.5      & 77.9  & 73.1 \\
TRILL top                       & 56.8      & 83.9  & 72.2 \\ 
TRILL layer 19                  & \textbf{57.0}& \textbf{84.7}  & 74.7 \\ 
TRILL distilled                 & 56.8      & \textbf{84.7}  & 74.7 \\ 
\bottomrule
\end{tabular}
\label{tab:results-intra}
\end{table}


\subsubsection{Fine-tuning}

Shallow models trained on top of frozen embeddings sometimes do not have enough degrees of freedom to adapt to mismatches between the train and target distributions, so we also experimented with fine-tuning the entire model end-to-end. For benchmarks tasks with relatively small amounts of data, we applied early stopping.

We also experimented with intra-speaker fine-tuning. For this we used the CREMA-D dataset, with fixed train/dev/test splits of 80\%/10\%/10\%. We first fine-tuned a common model on the training partitions of all the speakers, then further fine-tuned on each speaker separately. 

\section{Results}

\begin{figure}
    \includegraphics[width=\columnwidth]{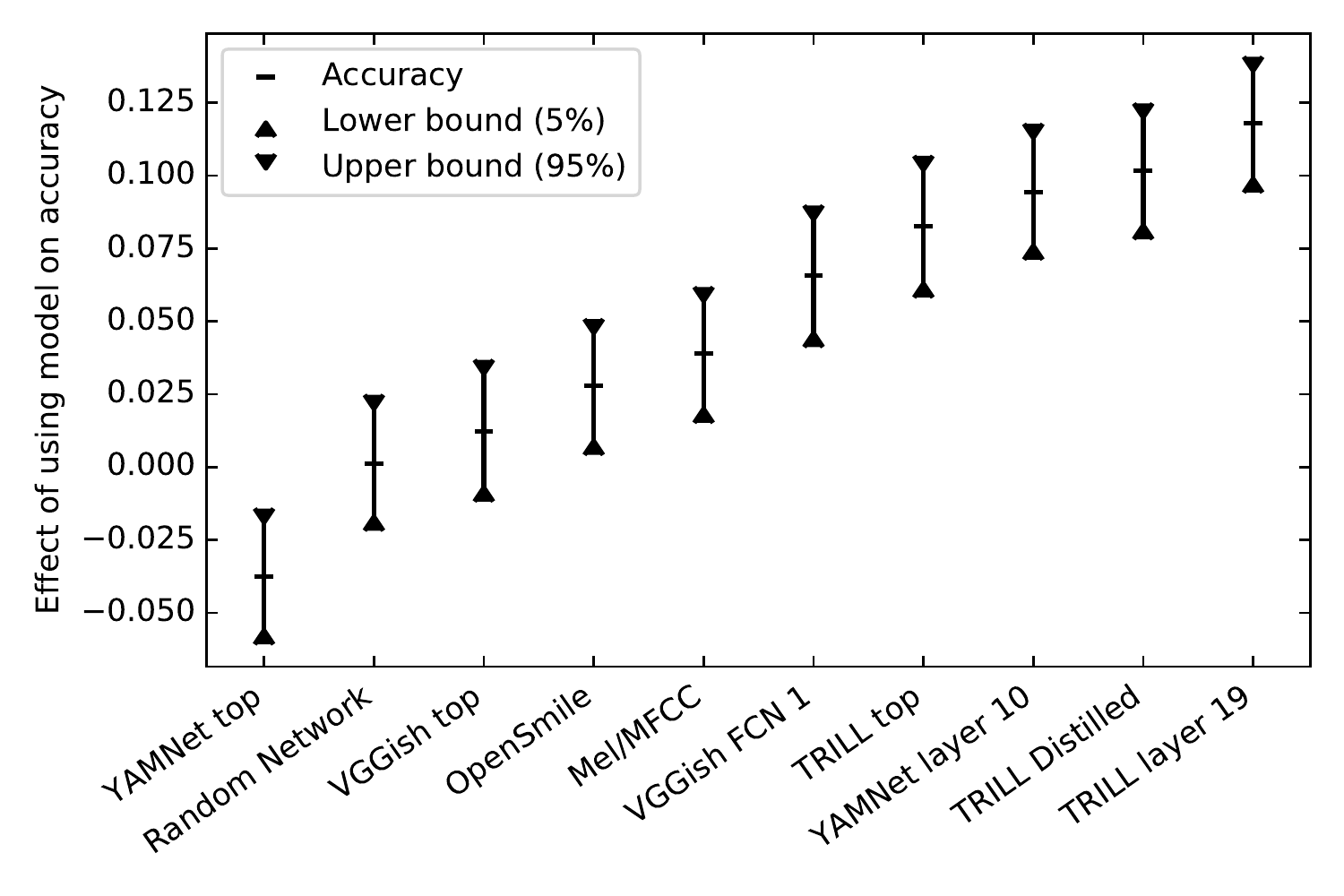}
    \vspace{-0.25cm}
     \caption{Effect of model on accuracy ($R^2=91\%$). A linear regression on the observed accuracies, with both the model and task as the explanatory variables. The effect a model has on the accuracy is the coefficient associated with the model in the regression. For a given task, when changing from one model to another, the resulting change in accuracy is expected to be the difference in $y$ values in this figure.}
    \label{fig:model_comparison}
\end{figure}
TRILL outperforms previously reported results on three of the six benchmark tasks, and is competitive with previous best on two of the three remaining tasks. On these two tasks, the best previously reported numbers use other modalities in addition to audio (visual or textual features) (Table \ref{tab:results-inter}). Of the representations we compared against, TRILL performed the best on five-of-six tasks (Table \ref{tab:results-inter}) and two-of-three intra-speaker tasks (Table \ref{tab:results-intra}). We successfully distilled TRILL to a much smaller model that can be used on a mobile device. The distilled model has no performance degradation on five-of-nine tasks, statistically insignificant degradation on one, and minor degradation on the remaining tasks. We compare across all tasks by fitting a linear regression on the observed accuracies, with both the model and task as the explanatory variables (see Figure \ref{fig:model_comparison}).

The distilled model results are presented in the one-before-last line of Table~\ref{tab:results-inter}. With the exception of VoxForge language identification and SAVEE speech emotion recognition, the reduction in model capacity and dimensionality has performance within the standard deviation of the larger embedding (variance calculated over 5-fold data splits). In the personalized tasks the quality of the distilled model is the same as the larger model.


\section{Analysis}

As can be seen in Table~\ref{tab:results-inter}, fine-tuning the final embedding gives a clear boost on most tasks and sets a new state-of-the-art in 3 out of the 6 datasets. This approach of learning a strong representation on a large benchmark and then fine-tuning it for the downstream task has been proven very effective in other domains \cite{zhai2019visual}, and in this paper we demonstrate the same is true in the speech domain.

An important observation of our research is that the effective representation learned by both YAMNet and TRILL is not at their final layer, but in fact in one of their intermediate layers. This intermediate layer must capture information which is later discarded in the final layers. The reasoning might be that when learning the triplet loss, the network might learn to discard properties which vary temporally such as tone of voice or semantic information, but this embedding is still learned in the intermediate layer. Another evidence for that can be seen in the results on the Speech Commands dataset, where the performance of the intermediate layers of all of the learned representations is much better than the performance of their respective top layer.


\begin{table}
\centering
\caption{Intra-speaker fine-tuning}
\begin{tabular}{ c c } \toprule
                                & CREMA-D \\
\midrule
TRILL frozen                    & 57.0 \\
TRILL global fine-tuning        & 69.6 \\
TRILL per-speaker fine-tuning   & 73.2 \\
\bottomrule
\end{tabular}
\label{tab:results-intra-finetune}
\end{table}

Table~\ref{tab:results-intra-finetune} shows that fine-tuning per speaker allows to further personalize to each speaker, and generally improves accuracy. When breaking down the performance impact on each speaker, we can see per-speaker fine-tuning improves accuracy for 31 speakers, is mostly unchanged for 49 speakers, and decreases for 12 speakers.

\section{Conclusions}

In this work, we explore the importance of clearly defining benchmarks when comparing representations of speech. We propose NOSS (Section \ref{sec:benchmark}) to help fairly compare non-semantic speech representations, and we introduce a sub-category of personalization tasks to help measure progress in the age of on-device computation.  We also demonstrate that TRILL (Section \ref{sec:triplet}), based on a self-supervised training criteria, simultaneously performs well on all benchmark tasks. We show that finetuning TRILL on a small amount of data outperforms or is competitive with almost all previously reported numbers for the NOSS tasks, and that TRILL is significantly better than other representations. Finally, we distill TRILL to be on-device with very little or no performance loss. The NOSS benchmark, the evaluation code, TRILL, and TRILL-distilled are all publicly available.

\bibliographystyle{IEEEtran}

\bibliography{mybib}


\end{document}